# Modeling a Grid-Connected PV/Battery Microgrid System with MPPT Controller

Genesis Alvarez[1], Hadis Moradi[1], Mathew Smith[2], and Ali Zilouchian[1]

[1]Florida Atlantic University, Boca Raton, FL, 33431, USA

{genesisalvar2013, hmoradi, zilouchi} @fau.edu

[2]IEEE Smart Village Volunteer, Piscataway, NJ, 08854, USA chemicalbull03@gmail.com

*Abstract* — This paper focuses on performance analyzing and dynamic modeling of the current grid-tied fixed array 6.84kW solar photovoltaic system located at Florida Atlantic University (FAU). A battery energy storage system is designed and applied to improve the systems' stability and reliability. An overview of the entire system and its PV module are presented. In sequel, the corresponding I-V and P-V curves are obtained using MATLAB-Simulink package. Actual data was collected and utilized for the modeling and simulation of the system. In addition, a grid-connected PV/Battery system with Maximum Power Point Tracking (MPPT) controller is modeled to analyze the system performance that has been evaluated under two different test conditions: (1) PV power production is higher than the load demand; (2) PV generated power is less than required load. A battery system has also been sized to provide smoothing services to this array. The simulation results show the effective of the proposed method. This system can be implemented in developing countries with similar weather conditions to Florida.

## I. INTRODUCTION

The massive increase in the global demand for energy is due to the industrial development, population growth and economic development. Many people in the world are currently experiencing dramatic shifts in lifestyle as their economies make the transition from subsistence to an industrial or service base. The largest raise in energy demand will take place in developing countries where the proportion of global energy consumption is expected to increase from 46 to 58 percent between 2004 and 2030 [1-6]. As environmental issues continue to progress due to conventional energy systems. Renewable energy sources such as photovoltaic (PV) is becoming a more promising source of energy for power generation. The amount of photovoltaic installations has exponentially grown, primarily due to the decrease in cost of solar panel systems [7-9]. The electric grid is an unlimited amount of energy supply. A photovoltaic system can either be a stand-alone system or a grid-tied system. In both systems a battery storage unit is often essential to the entire system. Therefore, battery management is significant to the functional life of the battery bank. The battery bank can also be implemented in smoothing the PV output power fluctuations.

The battery bank stores energy produced by PV panels. If the ramp rate of the PV is greater than the load can tolerate, then the battery bank serves to produce or consume power in order to smooth the output. One of two methods of charging the battery bank includes using the PV panels to charge the batteries during peak sun hours. In the occasion of cloudy weather, the lack of solar radiation energy doesn't permit the battery bank to recharge [10]. In a grid-tied PV system, distribution lines is a backup power source. Specific battery requirements in designing a PV array include battery charging losses, load demand, discharge rate, battery size, and storage temperature. Weather conditions, lower solar irradiation and higher temperatures are known causes for lower energy efficiency production by the PV array. However, there is an on-going research pertaining to voltage control of a PV array with maximum power point tracking (MPPT) and battery storage [11]. The MPPT algorithm supports sustainable efficiency by dynamically adjusting the voltage to ensure power optimization [12].

A Battery Energy Storage System (BESS) will be beneficial not only on a daily saving but reducing the PV output power fluctuations. The aim of this paper is to present the performance evaluation of the FAU fixed grid-connected PV array with a MPPT algorithm and a BESS with a time moving average algorithm to reduce voltage sag.

## II. SYSTEM OVERVIEW

In this paper. The performance of a solar unit located at FAU in Boca Raton, Florida as shown in Fig.1 is analyzed. The fixed solar array is installed on an East-West axis with a 23° inclination angle. The 12x2 PV array is connected to a Sunny Boy 7000US-12 inverter and then ties to the Florida Power and Light (FPL) utility grid. Features within this inverter are arc-fault circuit interrupter, measuring channel and MPPT algorithm. It is proposed to install a sealed lithium-ion battery storage system to the current grid-connected array.

## III. PV CELL MODELING

A p-n junction fabricated in a layer of a semiconductor forms a photovoltaic cell structure. The ideal solar cell is a semiconductor diode connected in parallel to a current source with series resistance, and parallel resistance as shown in Fig. 2 [13].

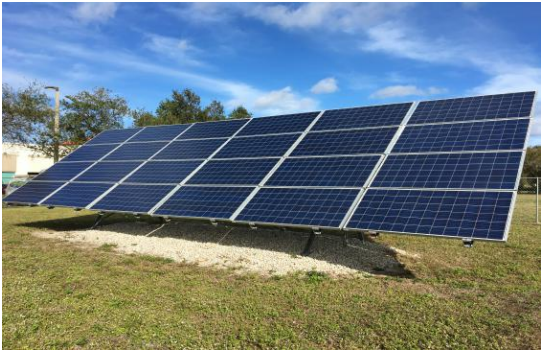

Fig. 1. Photovoltaic unit installed at FAU Boca campus

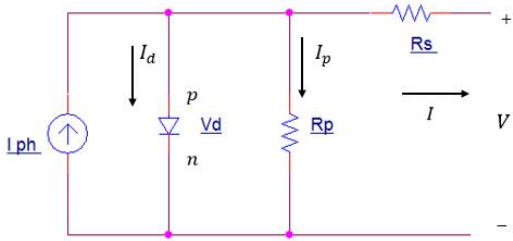

Fig. 2. Equivalent circuit of a PV cell with single-diode model

$$I = I_{ph} - I_0[e^{\frac{q(V+R_sI)}{nkT}} - 1] - \frac{V + R_sI}{R_{ph}} \quad (1)$$

$$V_d = V + IR_s \quad (2)$$

A simple equivalent photovoltaic cell circuit model includes a diode in parallel with a current source and a resistor as shown in Fig. 2. In the characteristic equation, $I_{ph}$ represents a current source created by sunlight known as a photocurrent (A), $I_o$ is the saturated current, $V$ denotes an output voltage (V), $V_d$ is the diode voltage, q is an electron charge of $1.6 \times 10^{-19}$, K is Boltzann constant of $1.38 \times 10^{-23}$, $R_s$ is the resistance connected in series, $R_s$ denotes a series resistance (Ω), $R_p$ is the shunt resistance across the diode (Ω) which deals with shading problems, $T$ is consider the cell temperature, $n$ is a deviation factor from the ideal p–n junction diode. Two important parameters within photovoltaics are the short-circuit current $I_{sc}$ and the open-circuit voltage $V_{oc}$ [6].

The fixed solar array located at FAU is composed of Multicrystalline Silicon cells. Parameters provided by Trina Solar indicate that each solar cell has a nominal operating cell temperature (NOCT) of 45 C°. After substituting and rearranging (1), the simplified equation is as follows:

$$I = I_{ph} - I_0(e^{36.44V_d} - 1) \quad (3)$$

## IV. SOLAR PV ARRAY MODEL

Although, there are varies types of solar cell materials, silicon is currently a dominant material utilize in solar cells due to its scalability, momentum and efficiency in light absorption [5]. The fixed PV array has a minimum array size of 6.84kW. This PV array consists of 12 modules per string connected in series and 2 strings in parallel resulting to 24 modules per array. Each 285W solar panel contains 72 PV cells connected in series as shown in Fig. 3. The parameters of the solar module under study are shown in Table I.

TABLE I
MODULE MANUFACTURE SPECIFICATION

| Electrical specs | Value |
|---|---|
| Module efficiency-$\eta_m$ | 14.7 % |
| Power output tolerance | 0/+3 % |
| Maximum power voltage | 35.6 V |
| Maximum power current | 8.02 A |
| Open circuit voltage-$V_{oc}$ | 44.7 V |
| Short circuit current | 8.5 A |
| Peak power watts-$P_{max}$ | 285 W |

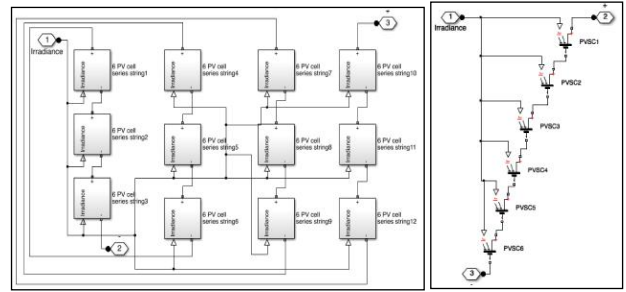

Fig. 3. (Right) Solar panel block diagram, (left) Six solar cells in series located in each subsystem

A dynamic model of the studied multicrystalline silicon cell Trina TSM 285 PA14 is presented in Fig. 4. In the I-V curve and P-V curves of the panel under 400, 600, 800 and 1000 W/m$^2$ solar radiation are presented is in Fig. 5.

The MPPT controller tracks the output power of the PV array in real time through the adjustments of the DC-DC converter. The MPP is dependent on ambient temperature and insolation shifts [14]. An algorithm is used to calculate the voltage and current of the PV output. There are varies types of algorithms; the two widely applied methods in MPPT are Perturbation and observation (P&O) [15] and incremental conductance (INC). Both methods regulate the PV voltage base on power delivered. In the P&O method since the voltage is increased the power delivered also increases, the voltage will continue to increase until the maximum power point (MPP) as shown in Fig. 6.

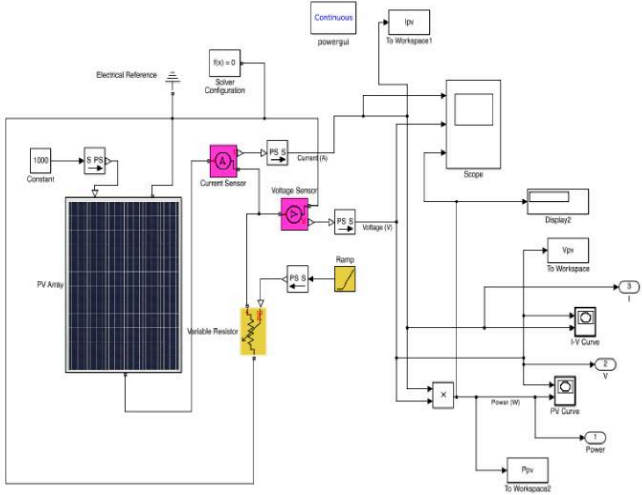

Fig. 4. The modeling of a 285W Trina solar panel

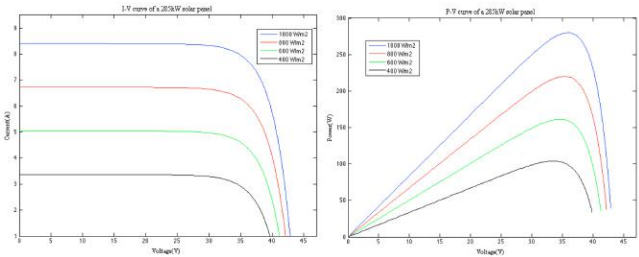

Fig. 5. I-V and P-V curves of a 285W solar model

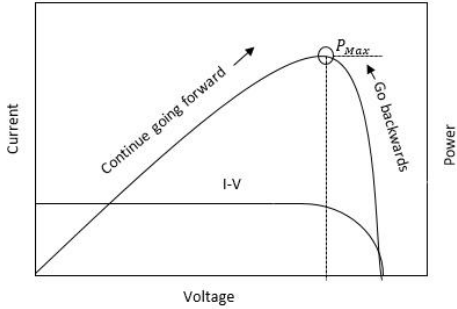

Fig. 6. Perturb - and - observe approach [7]

The incremental conductance algorithm (INC) is founded upon the fact that the MPP can be evaluated when the power versus voltage curve is equal to 0. Therefore it is determined that $dP/dV$ is equal to zero. Since $P=VI$, the MPP can be manipulated as shown in (4)-(7). The operating current and voltage is resembled in equation (1-4) as I and V. The ratio $\Delta I/\Delta V$ is consider as the incremental conductance.

$$\frac{dP}{dV} = I\frac{dV}{dV} + V\frac{dI}{dV} = I + V\frac{dI}{dV} \approx I + V\frac{\Delta I}{\Delta V} \quad (4)$$

$$\text{At the MPP: } \frac{\Delta I}{\Delta V} = -\frac{1}{V} \quad (5)$$

$$\text{The left of MPP: } \frac{\Delta I}{\Delta V} = -\frac{1}{V} \quad (6)$$

$$\text{The right of MPP: } \frac{\Delta I}{\Delta V} = -\frac{1}{V} \quad (7)$$

The INC algorithm measures the fixed incremental change of the solar panel conductance and compares it against the instantaneous conductance, which is known to be the ratio of $I/V$. Once the to are compared, the position of the operating current and voltage changes in reference to the MPP.

## V. BATTERY STORAGE DESIGNING

The capacity and life of the battery is dependent on the rate of discharge, depth of discharge and temperature. In this particular project, lithium-ion phosphate batteries will be used due to its robust life cycle characteristics and safety record .BESS are utilized for peak shaving, peak shifting and smoothing. Smoothing is used to smooth the generated solar power fluctuation which occur during periods with transient cloud shadows on the PV array. Smoothing also regulates the battery state of charge (SOC) under proposed conditions. Moving average configuration is applied for this small-scale BEES based smoothing [16]. In this case a $\pm 700W$ ramp rate will trigger the BESS to output power for smoothing.

Many industries use the statistical software Minitab which some of the features includes basic statistics, measurement analysis, regression and graphics. This program was used to create the histogram graph in Fig.7 and Fig.8. A 4kW/2kWhr lithium-ion phosphate BESS has been designed according to Fig.7 and Fig.8. The historical worst case ramp rate was used for this design criteria. The service duty of the battery in this case was found to be 3.65 kW and 0.91 kWhr. The battery has to be ready to accept charge or discharge power so a resting state of charge should be somewhere near 50%. The battery should accept up to 1 kWhr or discharge 1 kWhr thus a 2 kWhr battery was selected.

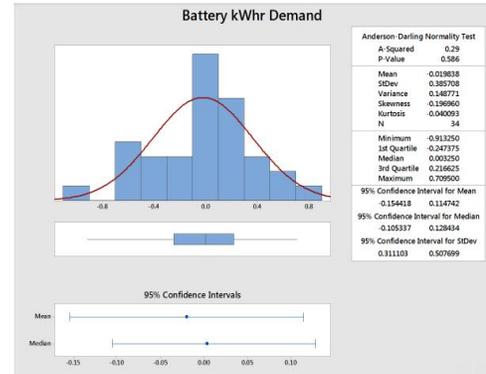

Fig. 7. Battery kWhr demand

The daily cycling on such a battery would be 5-10 partial cycles. Using vendor cycle life degradation curves it was found that cycling had less than 1% energy capacity degradation annually. A standard miner's rule fatigue calculation was used [17]. In this case, calendar degradation will dominate. The total energy capacity degradation rate will be approximately 2% annually. The battery size of 2kWhrs will support 8-10 years of the smoothing application under this scenario. Battery augmentations should be planned for year 10 to continue to serve this application. The worst-case scenario for voltage sag is during the summer in Florida. Voltage sag occurs in the occasion when the root-mean-square (rms) voltage drops below 90% nominal voltage, which causes a power disturbance. Depending on the duration of the voltage drop the voltage sag is distinguish with different titles. If the duration is 0.5-30 cycles this is consider an instantaneous sag, up to 3 seconds is consider a momentary and temporary sags the duration could go up to a minute [18].

Voltage sag is caused by abrupt increases in loads, short circuits and over-current condition, normally due to faults. In Fig. 9-11 the voltage sags are distinguished and the battery storage system is utilize to improve the effects voltage sags may have on the equipment.

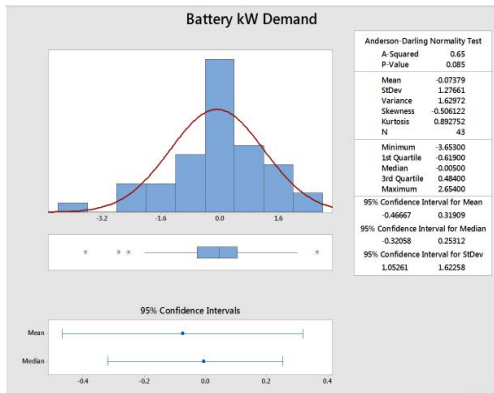

Fig. 8. Battery kW demand

Regarding the results, the battery storage smooths and shifts the system power output graph, which shows the effectiveness of BESS on system performance.

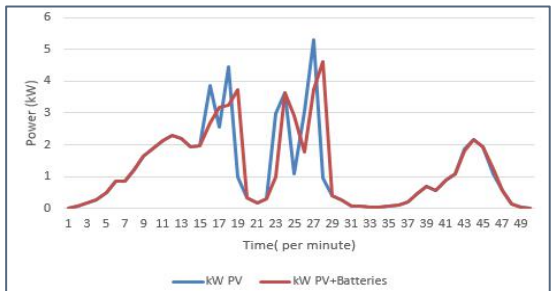

Fig. 9. PV output on June 07

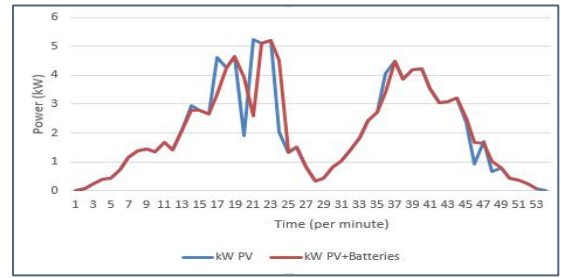

Fig. 10. PV output on June 26

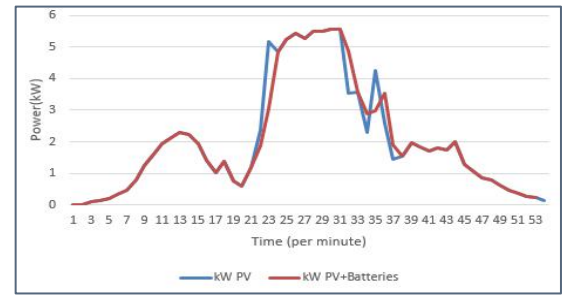

Fig. 11. PV output on August 26

VI. SIMULATION RESULTS

The integration of the PV and battery connected to the utility grid is simulated in MATLAB/Simulink using Simscape power systems toolbox as shown in Fig. 12.

The model composed of an array of PV panels connected to the DC/DC boost convertor, MPPT controller, a battery set connected to the bidirectional DC/DC converter, a 3-phase DC/AC inverter associated with 3-phase local load and the main grid. The output from the array system is provided as input to the boost converter to produce a regulated output of 24V. In sequel, the system can be used for charging the battery or to feed a 3-phase inverter. The initial battery state of charge (SOC) is assumed to be at 100% charge capacity.

Two cases are considered for analyzing the system performance under different test conditions. First, it is assumed that the 6840W PV solar system operates with 600 W/m$^2$ radiations [19]. The system is also tested under Standard Test Condition (STC) with radiation of 1000W/m$^2$ as the second case. The model parameters are adjustable to work with various solar radiations. The simulation results of the first and second case are shown in Fig.13 and Fig.14 respectively. The 3-phase load current is fixed because of the constant load at that time. Fig.13 shows the simulation results when $P_{pv}<P_{load}$. In this case, the battery is discharged to provide the additional needed power to load, and the DC current output of the BESS is indeed positive. The main grid also works normally to feed the load deficit. In second case, $P_{pv}>P_{load}$, therefore, the PV DC current output is higher, and the battery is in charging mode. Thus, the BESS current is negative with excess power to the utility grid. As the consequence of power feeding to the grid network, the current amplitude of both inverter and external grid are higher in comparison to the first case.

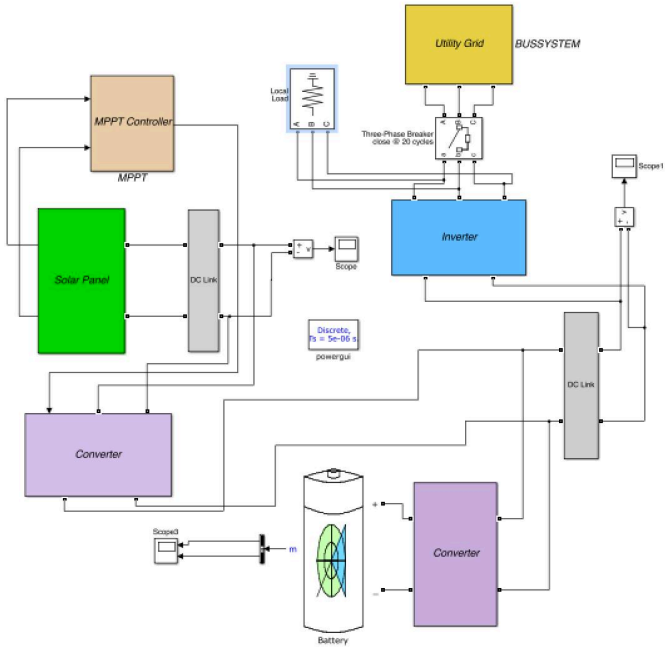

Fig. 12. Simulation model of a grid-tied PV solar system connected to battery storage

## VII. CONCLUSION

In this paper, different methods applied in MPPT were evaluated such as INC and P&O. A BESS was designed to smooth the PV array output. Based on the calculations, a lithium-ion battery with the capacity of 4kW was selected and integrated with the PV system. The results show that the PV/Battery dynamic model works properly and the system has reasonable reactions to the environmental and technical changes. In future work, other applications of battery storage will be studied such as peak shifting and voltage support. Also PV/BESS systems have the potential to be applied on residential rural areas in developing countries, where there is not much power consumption due to the absence of AC units, pool pumps and other appliances. In developing regions such as Africa, South East Asia, Latin America and the Middle East, 1,186 million people have no access to electricity. Applying renewable power allows energy service to be reliable, affordable and environmental friendly. Currently, rural committees are obligated to use polluting energy sources such as heating oil, wood and coal. Also, by improving the energy necessity, developing countries can achieve a long-term economic prosperity. This will reduce workload, increase jobs and improve education.

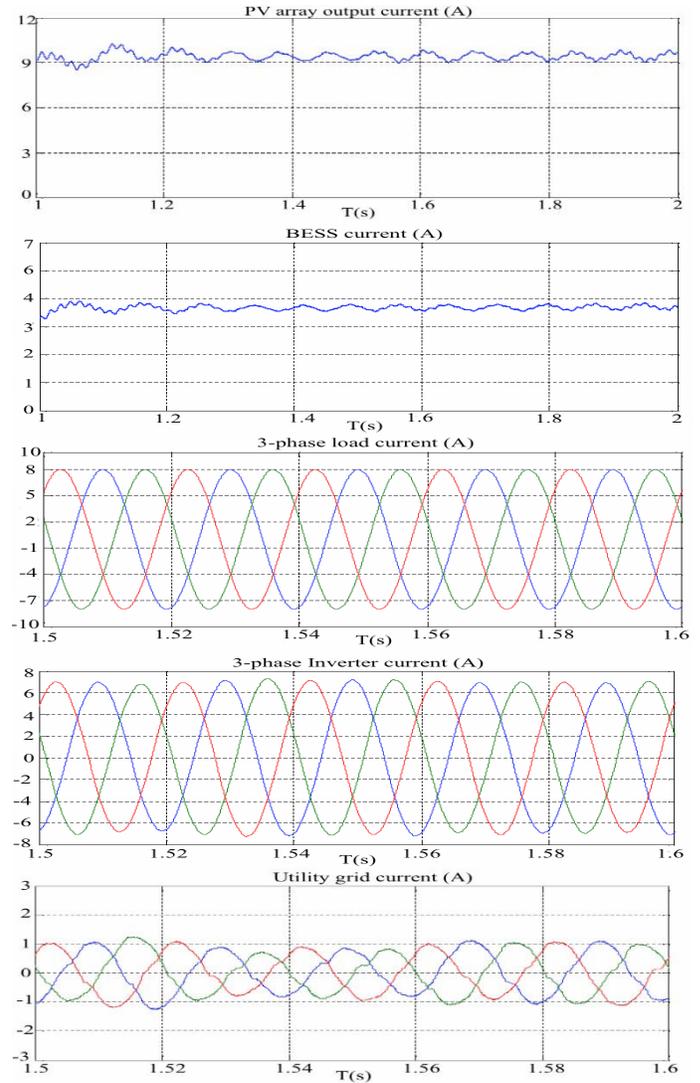

Fig. 13 First case simulation results

ACKNOWLEDGMENTS

The authors would like to thank Dr. Vichate Ungvichian for his valuable comments and also Florida Power and Light (FPL) Company for financial and technical supports through this project.ACKNOWLEDGMENTS

The authors would like to thank Dr. Vichate Ungvichian for his valuable comments and also Florida Power and Light (FPL) Company for financial and technical supports through this project.

## REFERENCES

[1] Y. Zhao and H. Khazaei, "An incentive compatible profit allocation mechanism for renewable energy aggregation," *2016 IEEE Power and Energy Society General Meeting (PESGM)*, pp. 1-5, Boston, MA, 2016.

[2] A. S. Mobarakeh, A. Rajabi-Ghahnavieh and A. Zahedian, "A game theoretic framework for DG optimal contract pricing," *IEEE PES ISGT Europe 2013*, pp. 1-5, Lyngby, 2013.

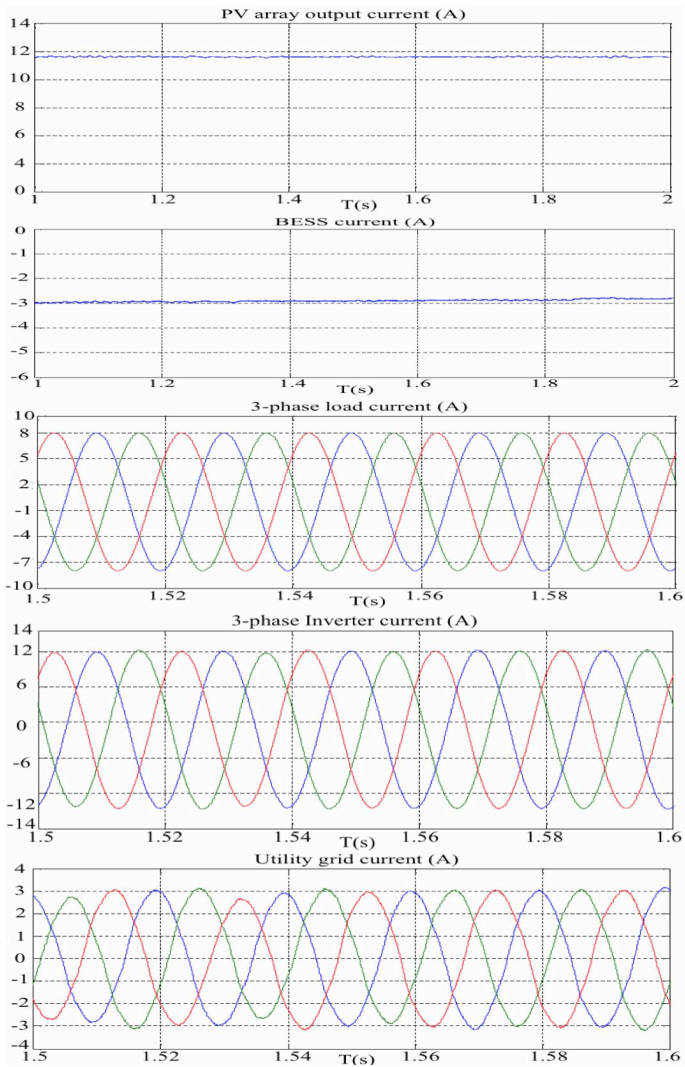

Fig. 14. Second case simulation results

[3] M. Farajollahi, M. Fotuhi-Firuzabad, A. Safdarian, "Deployment of Fault Indicator in Distribution Networks: A MIP-Based Approach," *IEEE Transactions on Smart Grid*, vol.PP, no.99, pp.1-9.

[4] M. E. Raoufat, A. Khayatian, A. Mojallal, "Performance Recovery of Voltage Source Converters With Application to Grid-Connected Fuel Cell DGs," *IEEE Transactions on Smart Grid*, vol.PP, no.99, pp.1-8.

[5] A. Shahsavari, A. Sadeghi-Mobarakeh, E. Stewart and H. Mohsenian-Rad, "Distribution grid reliability analysis considering regulation down load resources via micro-PMU data," *2016 IEEE International Conference on Smart Grid Communications (SmartGridComm)*, pp. 472-477, Sydney, NSW, 2016.

[6] H. R. Sadeghian and M. M. Ardehali, "A novel approach for optimal economic dispatch scheduling of integrated combined heat and power systems for maximum economic profit and minimum environmental emissions based on Benders decomposition, " *Energy*, vol. 102, pp. 10– 23, 2016.

[7] M. H. Athari, Z. Wang, and S.H. Elyas, "Time-Series Analysis of Photovoltaic Distributed Generation Impacts on a Local Distributed Network," *2017 IEEE PowerTech Conference*, pp. 1–6, 2017.

[8] H. Sadeghian, M. H. Athari and Z.Wang, "Optimized Solar Photovoltaic Generation in a Real Local Distribution Network," *2017 IEEE Innovative Smart Grid Technologies Conference (ISGT)*, 2017.

[9] L. Cadavid, M. Jimenez and C. J. Franco, "Financial analysis of photovoltaic configurations for Colombian households," *IEEE Latin America Transactions*, vol. 13, no. 12, pp. 3832-3837, Dec. 2015.

[10] S. Duryea, S. Islam and W. Lawrance, "A battery management system for stand-alone photovoltaic energy systems," *IEEE Industry Applications Magazine*, vol.7, no. 3, pp. 67-72, Jun 2001.

[11] E. B. Ssekulima and A. A. Hinai, "Coordinated voltage control of solar PV with MPPT and battery storage in grid-connected and microgrid modes," *18th Mediterranean Electrotechnical Conference (MELECON)*, pp. 1-6, Lemesos 2016.

[12] L. Tang, W. Xu, C. Zeng, J. Lv and J. He, "One novel variable step-size MPPT algorithm for photovoltaic power generation," *IECON 2012 - 38th Annual Conference on IEEE Industrial Electronics Society*, pp. 5750-5755, Montreal, QC 2012.

[13] J. L. Gray, *The Physics of the Solar Cell*, in Handbook of Photovoltaic Science and Engineering, eds A. Luque, S. Hegedus, Manchester, UK: John Wiley & Sons, 2010.

[14] Renewable and Efficient Electric Power Systems, Gilbert M. Masters, Second Edition, ISBN: 978- 1-1181-4062-8, 2013, Wiley-IEEE Press

[15] M. A. G. de Brito, L. Galotto, L. P. Sampaio, G. d. A. e Melo and C. A. Canesin, "Evaluation of the Main MPPT Techniques for Photovoltaic Applications," in *IEEE Transactions on Industrial Electronics*, vol. 60, no. 3, pp. 1156-1167, March 2013.

[16] D. Shwetha and S. Ramya, "Comparison of smoothing techniques and recognition methods for online Kannada character recognition system," *2014 International Conference on Advances in Engineering & Technology Research (ICAETR - 2014)*, pp. 1-5, Unnao, 2014.

[17] M. Safari, M. Morcrette, A. Teyssot, and C. Delacourt, "Life-Prediction Methods for Lithium-Ion Batteries Derived from a Fatigue Approach: I. Introduction: Capacity-Loss Prediction Based on Damage Accumulation," *J. Electrochem. Soc.*, vol. 157, no. 6, pp. A713–A720, Jun. 2010.

[18] IEEE Guide for Voltage Sag Indices, in IEEE Std 1564-2014, pp.1-59, June 20, 2014.

[19] H. Moradi, A. Abtahi and R. Messenger, "Annual performance comparison between tracking and fixed photovoltaic arrays," *2016 IEEE 43rd Photovoltaic Specialists Conference (PVSC)*, pp. 3179-3183, Portland, OR, 2016.